\documentclass{article}
\usepackage[english]{babel}
\usepackage[usenames]{color}

\begin{document}

\title{Business Mereology: Imaginative Definitions of Insourcing and Outsourcing Transformations
\thanks{This work has been performed  in the context of the NWO project Symbiosis.
Thanks are due to Guus Delen (Verdonck, Klooster \& Associates, Zoetermeer, The Netherlands),
Karl de Leeuw (University of Amsterdam) and Arjen Sevenster (University of Amsterdam) for discussions concerning sourcing.}
}
\author{J.A. Bergstra$^{1,2}$\\S.F.M. van Vlijmen$^{1}$
\\[2ex]
{\small\begin{tabular}{l}
  ${}^1$ Section Theory of Computer Science,
  Informatics Institute,\\
  Faculty of Science,
  University of Amsterdam.\\
  ${}^2$ Department of Computer Science,
 Swansea  University.
  \end{tabular}
}
}
\date{}

\maketitle

\begin{abstract}
\noindent Outsourcing, the passing on of tasks by organizations to other organizations, often including the personnel and means to perform these tasks, has become an important IT-business strategy over the past decades.

We investigate imaginative definitions for outsourcing relations and outsourcing transformations. Abstract models of an extreme and unrealistic simplicity are considered in order to investigate possible definitions of outsourcing. Rather than covering all relevant practical cases an imaginative definition of a concept provides obvious cases of its instantiation from which more refined or liberal definitions may be derived.

A definition of outsourcing induces to a complementary definition of insourcing. Outsourcing and insourcing have more complex variations in which multiple parties are involved. All of these terms both refer to state  transformations and to state descriptions pertaining to the state obtained after such transformations. We make an attempt to disambiguate the terminology in that respect and we make an attempt to characterize the general concept of sourcing which captures some representative cases.

Because mereology is the most general theory of parthood relations we coin \emph{business mereology} as the general theory in business studies which concerns the full variety of sourcing relations and transformations.
\end{abstract}

\section{Introduction}\label{sec:Intro}
In Delen \cite{Delen2005,Delen2007} a survey is given of definitions of outsourcing, insourcing, outtasking, intasking, follow-up sourcing, backsourcing, greenfield outsourcing and greenfield insourcing.\footnote{As early references concerning outsourcing Delen mentions: \cite{KernWillcocks1999}, \cite{Lacity1993}, \cite{LohVenkatraman1992}.} Recently multiple outsourcing has become prominent and geographical aspects have
become crucial.\footnote{In \cite{Beulen2000} multiple outsourcing is subsumed under outsourcing.}
For instance, offshoring as a specific instance of outsourcing where the insourcer is located `offshore', which means quite distantly, not only in kilometers but also in terms of native language and, more importantly, business culture.\footnote{Offshoring may be driven by: a quest for cheaper labor, by a quest for more disciplined workforce, by a quest for better qualified personnel or by a mere lack of
qualified personnel. Quite different motives are conceivable, however. Offshore insourcers (service providers) may be very significant operators who are better able to keep up with international competition than any potential local providers. Yet another motive is that offshore insourcers may force an outsourcer into intended but problematic changes of its internal processes more quickly and effectively than any local insourcer would ever achieve.} Although outsourcing and insourcing often directly or indirectly concern or comprise IT, we understand these transformations at a general level, so including applications outside IT.

It is customary to view outsourcing and insourcing as a topic in sourcing. But we will not generalize the theme to sourcing before a definition of outsourcing, which seems to be the oldest and most classical of all sourcing transformations, has been worked out. At the same time we assume that insourcing is complementary to outsourcing to a high degree and that definitions provided for outsourcing naturally lead to similar definitions for insourcing. Whether or not the entire subject can be treated more symmetrically in terms of sourcing remains to be seen.

In any case, unlike outsourcing, sourcing can not plausibly be understood as a transformation and for that reason a theory that takes sourcing as its primitive instead of insourcing and outsourcing must take a different form. Throughout the text occurrences of outsourcing may often be generalized into insourcing or outsourcing, or even to sourcing once one has made up one's mind on the meaning of sourcing. Our overriding objective has been, however, to clarify at least one form of activities subsumed under sourcing, viz.\ outsourcing in the clearest possible way. We cannot at the same time provide definitions of specific concepts and write at the most general level of abstraction.

It appears that a workable though informal view on outsourcing is as follows. Companies, enterprizes, and organizations  all combine human activity, titles of ownership, artifacts, infrastructure, and structures of command and control into units with a more or less visible identity. Such units have an internal structure and in some cases:
\begin{quote}
 parts of a unit may be removed from that unit (the outsourcing unit) and incorporated in some other unit (the insourcing unit) while the working of the ensemble of units as seen from some distance remains more or less stable.
\end{quote}

\noindent This formulation is close to accepted definitions of outsourcing, a possible difference being that we might have abstracted stronger from business rationales and management instruments used (see, e.g., \cite{Delen2005}). It seems to characterize (define) outsourcing and insourcing as mechanisms in a satisfactory abstract fashion.

However, there are some questions which cast doubt on the suggestion that outsourcing can be explained with such definitions. We'll digress on these questions in Section~\ref{questions}. The motive of this paper is then to make a step towards a definition that we hope will be a starting point to answer these questions. So, we'll not try to answer them all, we rather take them as a justifying context to set-out for our approach, which we'll describe in Section~\ref{approach}. Given this footing we develop in Section~\ref{nconcepts} a set of basic terminology and definitions. These are basic in the sense that the definition of outsourcing is postponed to Section \ref{outsourcingdefined}. Before reaching that point we present ways to quantify and qualify a sourcing situation, i.e., we seek for forms of assessment that may be relevant from a business perspective, in Section \ref{qualiquanti}. A conclusion follows in Section \ref{conclusie}.

\tableofcontents

\section{Contemplating sourcing}\label{questions}
The following survey lists a sample of questions which can hardly be avoided if one contemplates the concept of out/insourcing.\footnote{We use `out/insourcing' to refer to the combination of outsourcing and insourcing both understood in the narrow sense which will be elaborated in the following sections.} There are some questions about the relationships between notions. Then, there are several questions that concern the parthood relations which are assumed between units in order to explain out/insourcing. Other questions stem from observed ambiguity in the current terminology and from the consequential lack of a clear approach when one sets out to theorize about out/insourcing.

\subsection{Relative meaning of notions}\label{relations-notions}
The following questions spring to mind when one contemplates the World Class IT-sour\-cing\-cycle \cite{Delen2000} or the `sourcing life cycle' as defined by \textit{Platform Outsourcing Nederland} \cite{PON2006}.
\begin{enumerate}
\item Is it essential to distinguish decision making from other work when defining outsourcing? Can outsourcing take place in a non-deliberate way, or must it necessarily be carried out on the basis of a preconceived plan? If so, is the planning process at hand sequential (see \cite{Simon1959})?
\item Are outsourcing and insourcing complementary concepts that need to be defined in one go, or can, say, outsourcing be conceptually fixed while leaving room for variation in the meaning of insourcing?
    This question can be made more specific in several ways. For instance, is it necessarily the case that if unit $A$ outsources source $P$ to unit  $X$, then $X$ provides (after insourcing $P$) additional services to $A$ or can it be the other way around that $X$ provides fewer services to $A$? This seems to be possible if before transformation all or most services provided by $P$ are in fact being delivered to a third unit $C$ while $P$ makes use of services provided by $X$, in which case after the transformation $A$ and $X$ may not even interact anymore. One might claim that in this situation $P$ hardly constitutes a source for $A$, but that requires rather precise definitions as $P$ (before transformation) is a source inside $A$ and under control of $A$.
\item If backsourcing denotes the inverse transformation of outsourcing then it is unclear whether after outsourcing a state is obtained from which the complementary backsourcing is somehow permanently enabled. In other words, is the history of organizational change frozen in such a way that undoing this particular (outsourcing) transformation has a preferred status? Related questions are:
\begin{itemize}
\item If one adopts the `sale and lease-back' view of outsourcing, is the sale (as well as the lease) undone when follow-up sourcing takes place with another partner (from the outsourcer's perspective)?
\item Is outsourcing necessarily followed by follow-up sourcing after some time. If not, what kind of transformation has implicitly taken place?
\item Is backsourcing a special case of follow-up sourcing?
\end{itemize}
\end{enumerate}

\subsection{Ambiguity in terminology}
We hold that outsourcing is an extremely ambiguous term and this ambiguity may be considered a weakness of the conceptual framework of sourcing. The ambiguity is that out/insourcing serves as a state transition qualification and it simultaneously serves as a qualification for a stable operation or state. If $A$ is said to be outsourcing $P$ one simply cannot tell if the act of outsourcing is ahead or if it has been completed already.

The ambiguity of the term outsourcing can be compared with the ambiguity of the term diving, which also combines the reference to a visible and audible splash taking place in space and time with the reference to a less noticeable state of being submerged. Interestingly the diving state forecasts its termination. For fish who will stay submerged the term diving is uncommon.

Perhaps this fundamental ambiguity of the term outsourcing at the same time constitutes its strength. Nevertheless, we will propose to make a choice. More specifically we propose to use outsourcing as a reference to (qualification of) a state transformation rather than as a reference to a state. The argument is that given some sourcing equilibrium (defined in Section \ref{sourcing-equilibrium}) with a number of participating units, and in the absence of historical information, it is a subjective judgement to state that a certain unit is outsourcing some activity for say $70\%$. Making such a judgement requires some sort of reference model.\footnote{The data collected in the Symbiosis project, which takes a close look at around 25 outsourcing cases, suggests the tentative conclusion that this historical information is easily, and indeed often, lost. Published material on Symbiosis is forthcoming at the time of writing (December 2010).} For the state transformation interpretation of outsourcing no general framework is needed to understand its meaning, provided one is able to point out in functional terms which activities are subject to outsourcing.

Only if terms can be used systematically with a fixed meaning, further complex forms of usage can be devised. Thus we claim that disambiguation precedes complex usage. It may be so that for management consultants the ambiguity of the term is a strategic advantage. That might change, however, if more complex outsourcing transformations need to be contemplated.

Another way to deal with the ambiguity of outsourcing has been proposed in \cite{WibbelsmanMaiero1994} where a continuum of meanings for the word is laid down. This continuum is explained in detail by Michel Troost in \cite{Troost2009} which provides an unusually comprehensive survey of the IT-sourcing literature. At the same time Troost exemplifies the ambiguous use of the term outsourcing. He cannot be blamed for these semantic ambiguities of course, these emerge from existing literature, but the text of \cite{Troost2009} seems to have been rendered more complex than needed because of these ambiguities.

\subsection{Parthood issues}
Parthood is studied in its most general form in mereology, a subject that is usually considered to be a part of philosophy. In \cite{Varzi2009} one finds the principles of mereology set out in a concise and comprehensible way. We'll get back to
mereology in more detail in Section \ref{approach}.
\begin{enumerate}
\item What structure theory of units  is needed in order to obtain a meaningful picture of an outsourcing
transformation. Is there a notion of part or component that comprises states, tools, activities and contracts at the same time?
\item What makes units different? In the case of outsourcing, is overlap permitted or underlap (see \cite{Varzi2009} for these terms).
If the insourcing unit is created as the result of outsourcing a part of the outsourcing unit, what justifies the assertion that this is a new unit? In particular, why does the resulting cooperation between outsourcing unit and insourcing unit after the transformation not qualify them for being considered a single unit? More specifically:
\begin{itemize}
\item Are units controlled as if they execute a control code \cite{Bergstra2010b,BergstraMiddelburg2009,Janlert2008}
or is the assumption
of centralized and codified unit control immaterial for definitions of outsourcing transformations?
\item Are different units interacting in qualitatively different ways from the interaction between different parts within the same unit?
\item Is unit identity a matter of structural properties of  units, or are units exclusively (or primarily) determined by means of formal regulations and restrictions issued by some higher authority?\footnote{In the case of money the point of view that money
can only be defined in terms of regulations and restrictions has emerged in the BFH school of
economic thinking (based on ideas of Fisher Black, Eugene Fama and Robert Hall). See also \cite{Bergstra2010a}.}
\end{itemize}
\end{enumerate}

\subsection{Behavioral issues and modelling}
Although the paraphrased definition in the introduction indicates that `seen from some distance' the combination of insourcing unit and outsourcing unit `remains more or less stable', there is no explanation of what difference may result as the effect of outsourcing. Which theory of behavior is needed (or helpful) to explain what remains the same and what is modified? Different theories may come into play at different
levels of abstraction.
\begin{enumerate}
\item Is process algebra \cite{BaetenBastenReniers2009} applicable or is a far higher level of abstraction
needed? And if a higher level of abstraction is required is
thread algebra, an abstract process type set out in detail in \cite{BergstraMiddelburg2007}, sufficient for describing the required form of concurrency in this case? And in case branching time is considered  a hindrance rather than a help (for specifying the concurrency between insourcing unit and outsourcing unit) are the progression rings of \cite{BergstraPonse2009} more useful?
\item A money stream between insourcing unit and outsourcing unit is to be expected. Is a theory of money presupposed or can
outsourcing be conceptualized without any mention of financial mechanisms?
We mention \cite{BergstraMiddelburg2009b} as an approach to formalizing financial streams.
\item Concurrent units may have different roles. Is an insourcing unit after the outsourcing transformation a service provider to
such an extent that the concept of a
service needs to be defined first before any attempt to define outsourcing related concepts?
\item Specific roles of units may be too complex to be of use for definitions of out/insourcing. Indeed, if the concept of a service is too involved to serve as a prerequisite of a definition of outsourcing, can it be replaced by making use of the notion of a promise? See \cite{Burgess2007}.
\end{enumerate}

\section{Approach}\label{approach}
New definitions for outsourcing are of course primarily useful if that leads to new or improved applications. But convincing though indirect arguments can be provided for new work on definitions in the area of sourcing. Our incentive to suggest a new definition stems from observing some problematic conceptual weakness in the current usage of terminology -- see the previous section -- rather than from the contemplation of an challenging but yet unsolved sourcing issue.

In this section we first pick a name for the conceptual framework of outsourcing we intend to develop, see Section~\ref{businessmereologyenAD}. Next, in Section~\ref{imaginativedefs}, we discuss the `tool' for this project: the imaginative definition.

\subsection{Business mereology and aggregation dynamics}\label{businessmereologyenAD}
Parthood is the subject of mereology, a discipline that is usually considered to be a part of philosophy. We refer to \cite{Varzi2009} for an overview of mereology.

It makes sense to name our approach because we intend to set-up a line of thinking that departs from the familiar successful engineering, goal and application oriented style that is, for good reasons, prominent in many disciplines including IT-outsourcing today. We will use {\em business mereology} for this purpose. Business mereology deals with the static structure of units and their parts. We use `business mereology' rather than `unit mereology' to indicate a focus on units in which human activity is prominent.

One may think of many other names for the style of enquiry in the field of outsourcing as we try to put forward here. We discuss one candidates because this figures in an earlier joint publication \cite{BergstraVanVlijmen1998}: aggregation dynamics. There, aggregation dynamics is understood as a general systems theory that studies the behavior and development of aggregates of `units', which can be aggregates themselves of course. The rules that prescribe what belongs to a certain unit are extremely liberal relative to commonsense ordering as, for instance, found in biological taxonomies or business organization charts. The central question is what kind of transformation principles of matter, information and energy equilibria exist. In particular, what types of transformation are universal in the sense that they are found at many levels of complexity.

Aggregation dynamics is also a phrase which is used in diverse areas including fluid dynamics, colloidal chemistry and biophysics \cite{aggdyn-colloids}. It has been coined for use in self-organization \cite{agdyn-overlaynetworks} and also for use in ethology \cite{aggdyn-ecosystems}. However, an elaborate definition of the term does not seem to be used in these disciplines, it's merely a descriptive designation.

Aggregation dynamics may serve as a counterpart of business mereology by having a primary focus on the dynamics and evolutionary forces that drive the aggregation and disaggregation of packaged human activities with a business (company, enterprize) as a paradigmatic form of aggregation.

In the context of aggregation dynamics, insourcing and outsourcing are mere instances of transformation mechanisms. One may ask where in a ranking of transformation mechanisms that are explained by aggregation dynamics outsourcing and insourcing should be positioned. It seems obvious that economy of scale, specialization of labour, market power and the need to produce increasingly complex artifacts each constitute more important drivers of aggregation. And anti-cartel policies constitute a most important mechanism for disaggregation. Shareholder value constitutes a driver for aggregation as well as for disaggregation though often of a lesser sophisticated form than outsourcing transformations. Insourcing and outsourcing seem to come on a third position only in this ranking.

Looked upon it in this way, aggregation dynamics is a super theory for business mereology, or business mereology is an instantiated specialized form of aggregation dynamics.

It may well be, however, that in terms of the frequency of occurrence, as opposed to the visibility per event, out/insourcing are most important mechanisms second to none in the above listing. Because out/insourcing events are often hidden from the stock market, as if these imply only minor internal changes of organizational structure, their impact is found only indirectly, at least in comparison to business takeovers, IPO's and company decompositions. Assuming some vague understanding of these forces aggregation dynamic oriented meta-questions about out/insourcing can be posed.

\begin{enumerate}
\item Can outsourcing increase the complexity of a collection of units. If not, are out/in\-sour\-cing transformations better perceived as simplifications and which complexity increasing transformations are to be found in aggregation dynamics for which out/insourcing transformations act as balancing simplifications?
\item Suppose each unit and for each unit all of its subunits are characterized by a family of arrows where each arrow is labeled with a service which is provided from source to target together with a quantification of the strength of the flow. Money streams are also depicted as arrows. Now a family of units may be considered a labeled graph (henceforth network) and outsourcing becomes a network transformation. Suppose one collects data about such aggregates in practice and measures actually occurring outsourcing activities. Is there any general statistics to be expected? For instance, is it conceivable that out/insourcing transformations have a tendency to develop a network towards a scale free structure \cite{Barabasi1999}? Or might it be the case that nodes act as attractors for other nodes with a comparable structure but with lesser weights?
\item Aggregation dynamics may provide a language for depicting in intuitively clear and convincing way the look and feel of transformational events. For instance, the phenomenal disappointment that is created if an old company fails, or the sense of pride that goes along with a new investment. Out/insourcing events have similar but perhaps weaker psychological effects which nonetheless need careful attention. This is not so clear from the available literature. Putting such effects in a proper perspective may be easier accomplished in a more general setting.
\end{enumerate}

\subsection{Imaginative definitions}\label{imaginativedefs}
Our objective is not to answer all the questions listed in the introduction. Rather we use this listing of questions as a context from which to acquire a sense of direction concerning the development of  imaginative definitions of outsourcing transformations and sourcing relations.

A description of imaginative definitions can be found in \cite{Bergstra2010a}. We refer to \cite{Middelburg2010a} and \cite{Middelburg2010b} as evidence that imaginative definitions may be missing for concepts which have been heavily used in computing practice such as `operating system' and `program testing'. We suggest \cite{BergstraMiddelburg2007} (multi-threading) and \cite{BergstraLoots2002} (instruction sequences) as examples of attempts to provide imaginative definitions.

A concept may be provided with a family of imaginative definitions which vary in level of abstraction as well as in their degree of refinement. Any imaginative definition of out/insourcing should enable one to work out reasonable answers to a number of the questions listed above. Such answers may not bring much in practice but they provide a complete and consistent story which is quite hard to obtain in the context of sourcing.

\subsubsection{A definition in stages}
For an imaginative definition of some concept it is needed to have a context of related concepts available. This `concept under definition' can be used already before its definition has been completed without creating confusion, provided it is made clear when and where in the text the final stage of a definitional effort is reached.

The simplest and initial stage of such definitions (now applied to the specific case of out/insourcing)  is to state that an outsourcing is an element of the sort of outsourcings. However circular this may seem to be, it sets a scene where surrounding sorts (categories) for neighboring notions can be set up. The definition of outsourcing is completed at some stage when a sufficient amount of information has been collected. Arguably the choice of this final stage of an imaginative definition is an arbitrary design decision to some extent. This differs significantly from the situation found in mathematics and logic where major definitions may become frozen for the duration of centuries.

For the development of additional theory, further refinement and the inclusion of additional information in a definition may be needed. In logic and mathematics that process of definition maintenance can be avoided in principle by introducing a focus on specific subclasses (subsorts) of a concept. In more fluid circumstances it should be admissible to maintain or upgrade a definition when additional insights are acquired. We will make use of this freedom for definitional maintenance ourselves in Section~\ref{outsourcingdefined}.

Indeed, although a specific definition of some concept prepares it for usage in theory development, letting the specific concept (e.g., given by means of an imaginative definition) stand for and even to some extent replace the informal concept can hardly
be justified. Thus, all we may possibly achieve in this work
is to provide a definition of outsourcing which can be referred to as `outsourcing as defined by ... in ...'. Moreover, to the extent that our definitions below have a non-trivial internal complexity, it is plausible that alternative definitions can be found. Indeed achieving clarity rather than universality is the objective of developing imaginative definitions.

\section{Neighboring concepts for outsourcing}\label{nconcepts}
Below we will have a focus on outsourcing because that is the most classical and well-known outsourcing transformation. Simplified (imaginative) definitions should provide insight in the presence, absence or relevance of symmetries, in role (insourcing versus outsourcing) and in time (outsourcing versus backsourcing).

The following notions will be further detailed in this section in preparation of definitions of outsourcing, which is our final objective:
\begin{itemize}
\item units, parts, sources and activities (\ref{units});
\item source types (\ref{sourcetypes});
\item sourcing equilibria (\ref{sourcing-equilibrium});
\item dimensions on which to qualify outsourcing transformations (\ref{qualificationDimensions});
\item titles to sources (\ref{catalogoftitles});
\item outsourcing transformations and the precondition thereof (\ref{outsourcingTransformation});
\item postcondition of outsourcing transformations (\ref{postcondition}).
\end{itemize}

\noindent After these neighboring concepts have been specified subsequent stages of the definition of outsourcing can be developed. This results in the definition of an outsourcing progression in Section~\ref{progression}: a sequence of steps that constitutes a transformation between two sourcing equilibria. Finally, we will identify outsourcing with carrying out an outsourcing progression, that is moving through a progression which effects an outsourcing transformation in Section~\ref{outsourcingdefined}. All key notions and some derived notions will be succinctly presented in Section~\ref{termsurvey}.

Referring to the issues raised in Section~\ref{relations-notions}, we assume that outsourcing and insourcing are complementary or reciprocal transformations.  However, it will appear that backsourcing can't always be used to indicate the reversal of an outsourcing transformation (see Section~\ref{postcondition}).

\subsection{Units, parts, sources and activities}\label{units}
Units have parts also called  subunits. These perform activities which combine into unit behavior. Parts make use of sources for carrying out their activities. `Source' is a basic concept that underlies the theory and practice of sourcing but as such it need not be defined.

What matters is that sources can be counted, are physical, can be consumed, but may also be permanent (tools), sources may or may not require an investment. Access to sources is structured by legal means as well as by physical means. In an imaginative definition sources may be imagined as points from which streams of services emerge. Reverse streams may be present as well, for instance financial streams or feedback streams, or streams that convey inputs which are upgraded by an activity carried out by the source. In any given context some classification of sources is assumed beforehand. Most sources are used exclusively by a single unit (and a single part of that unit) but shared usage is possible too.

Outsourcing should preferably be defined at a level of abstraction where other details do not matter. However, below we will discuss the concept of a source in some detail. The added value of being specific about sources is that one can investigate whether complete descriptions of specific outsourcing transformations can be given in principle. Furthermore the notion of a source depends on a business intuition which we intend to preserve for the time being. However, on the long run sourcing theory may become fully independent of its historical roots.

We leave untouched the question which descriptions of outsourcing transformations can be called complete. We only notice that completeness is a relative matter with respect to a level of abstraction which need not be fixed in advance.

\subsection{A survey of source types}\label{sourcetypes}
Although the concept of a source acts as a parameter in the theory of sourcing it may be informative to spell it out in more detail as argued above. Rather than defining it we will list a range of forms that sources can take without any claim of completeness.

\begin{description}
\item{\em Real estate}. Including land, mines, wells, buildings, roads, harbors, airports, railroads, data communication network cables,
and fixed position infrastructure.
\item{\em Vehicles}. Including ships, airplanes, cars, trains, cycles, satellites.
\item{\em Works of art and culture}. Including paintings, statues, movies, antiquities.
\item{\em Tools and equipment}. Including computers, laptops, mobile phones, tools for mechanical and chemical engineering, design
tools, music instruments.
\item{\em Control codes}. Including computer software used to control real estate components, vehicles and tools and equipment, provided it
can be handled (bought, sold, stored, repaired, modified, maintained, reverse engineered) independently from its target entity.
\item{\em Information bases}. Including proprietary data collections that support business operations. Included are as well access rights to specific technical and scientific journals, market projections, statistics.
\item{\em Persons}. Including the human capital embodied in a workforce, the expertise available via friendly business connections.
\item{\em Intellectual Property Rights (IPR)}. Including proprietary information on designs, descriptions, patented inventions, copyrighted texts, brand names.
\item{\em Knowledge}. Knowledge includes tacit as well as explicit knowledge of personnel about the trade and craftsmanship of a unit but also about other sources and about business processes, as well as about other units and their past behavior. Dedicated ontologies constitute knowledge. There is a continuous transition from knowledge to information bases.
\item{\em Contracts and goodwill}. Including the portfolio of outstanding contracts and work in progress,
the connections with customer representatives, a quantified reputation.
\item{\em Finance}. Credit lines, financial holdings, equity, insurances.
\end{description}
In this listing the vast information freely available on the internet does not feature as a source and neither does the oxygen in the air. That may be wrong but such resources\footnote{We do not distinguish between {\em source} and {\em resource}, but we use the term source because that has become customary in sourcing theory.} seem not be covered by outsourcing deals.

Arguably a core competence is also a source because: it's preservation influences business decisions and it contributes to the binding between a unit and its employees. Moreover, it may make the unit an interesting business partner because other units
might want to tap the core competence. The business intuition is that by definition a core competence cannot be outsourced because by doing so the outsourcing unit changes its identity and this identity should be kept. However, from our point of view, in a sourcing process some sources are moved out and others are not. We rather consider a core competence as an element of an emergent valuation of a such a process. The decision to outsource $S$ instead of $P$ because the latter is considered a core competence can turn out to be false in retrospect.

\subsection{Sourcing equilibrium}\label{sourcing-equilibrium}
We take from economics the term equilibrium instead of state. A {\em sourcing equilibrium} refers to a fixed assignment of sources to all (parts of) units. That is the totality of all units that make use of a collection of sources in a certain pattern. The relevance of the notion of an equilibrium is that it refers to a state and at the same time accommodates the normal functioning of the whole system.

In a sourcing equilibrium the sourcing connections stay unchanged. The activity of various units may change, however. The usage made of specific sources may also gradually change as well. In any case, a sourcing equilibrium is dynamic in the sense that it allows for ongoing activity, just like an economic equilibrium which is a dynamic state in which price formation has become stable and all economic activity proceeds at a relatively fixed rate.

Outsourcing and insourcing both denote a progression of steps which lead to a new sourcing equilibrium. This specification of outsourcing and insourcing can be made more specific of course, we will do so below. But the point is that we conceive of both as transformations of a sourcing equilibrium rather than as qualifications of a sourcing equilibrium. We do not introduce a name for the sourcing equilibrium that results after such a transformation. One may speak of the transformed equilibrium after an outsourcing transformation.

In order to provide an imaginative definition of outsourcing we need to capture the dynamics of the transformation involved. Rather than describing or defining such transformations from their constituent parts, conditions are defined about the state before and after the transformation in Section~\ref{outsourcingTransformation} and \ref{postcondition}. Before we can come-up with the condition we need to specify certain dimensions to which the conditions can refer. We'll discuss four dimensions in the next section and will elaborate on one of them -- titles to sources -- in the section thereafter (\ref{catalogoftitles}).

\subsection{Four qualification dimensions for outsourcing transformations}\label{qualificationDimensions}
Below we will denote a unit $U$ after transformation as $U_R$ where $R$ stands for `retained'. Assuming that an outsourcing transformation takes place between $A$ and $X$ with subunits $B$ and $Y$ respectively, then the following aspects or dimensions can be distinguished for further qualification of the outsourcing transformation.

\begin{description}
\item{\em 1. Title dimension.} The shift in titles to sources and the volume thereof. Two typical questions here are the following. First, are some sources of $B$ abandoned? That means that $Y$ after the transformation has no title to them. Second, are some sources new for $B_R$ as well
as for $Y_R$?\footnote{Orthodox definitions of outsourcing suggest that sources
should change title to the insourcer rather than be abandoned. Our definition of outsourcing is neutral
on that matter, and we hold that the degree to which sources are or are not abandoned
is a property (parameter) of any individual outsourcing transformation. This is comparable to the price of an item at which it is sold,
the sale taking place independently of the price which is being set during the sales progression.}
\item{\em 2. Contractual dimension.} Contracts define to what extent and for which period of time $Y_R$ will provide services that
$B$ used to provide before the outsourcing progression took place. Such contracts (SLA's are often parts of these contracts) will deal with pricing, quality of service,
responsibility for innovation and continuing focus on efficiency, handling of emergencies and reaction to adverse economic
conditions.\footnote{The conventional definitions of outsourcing suggest the need for these contracts, however, if $B$ provides services
only to customers of $A$ (outside $A$) then no SLA between $A$ and $X$ needs to be agreed. It may also be the case that neither $A$ nor $X$
feel an incentive to regulate service provision from $Y_R$ to $A$ (or its customers) by way of a specific contract. This is the case
 if $Y_R$ can
deliver a service (that had been provided by $B_R$ prior to the outsourcing transformation) whenever
$A$ asks for a service at known market prices, while $Y_R$ can cope with the loss of $A$ as a customer at any moment.}
\item{\em 3. Operational dimension.} New management and operational methods and structures
 may be needed for $B_R$ and for $Y_R$. So, the services that
are delivered in the new sourcing equilibrium may be produced in different ways.\footnote{It is plausible that $Y_R$ provides
similar services (to the one's $B$ was providing to its sister units in $A$ and to $A$'s customers)  to other customers than units or
customers of $A$. This may provide $X$ with an economy of scale out of reach for $A$ using $B$ and the original sourcing equilibrium.}
\item {\em 4. Competitive advantage dimension.}
We restate a proposition from \cite{Barney1999}, which constitutes a remarkably clear and concise insight expressed in sourcing terminology:
\begin{quote} A resource provides a sustainable competitive advantage if the following characteristics are present:
the resource is {\em valuable} and {\em rare} and it {\em cannot be imitated} and it is {\em non-substitutable}.
\end{quote}
The competitive advantage dimension expresses the extent to which a unit is able to increase its focus on sources that provide it
with a competitive advantage.
\end{description}

\subsection{A catalogue of titles to sources}\label{catalogoftitles}
The previous section introduced a set of outsourcing qualification dimensions. Titles to sources are in this set and in this section these are elaborated. In Sections \ref{outsourcingTransformation} and \ref{postcondition} these titels to sources will be used in the specification of a pre- and postcondition of the outsourcing transformation as understood by us.

Again, assume that unit $A$ has subunit $B$. $B$ performs some services towards other (sub)units of $A$ and perhaps also to units external to $A$. In order to provide these services $B$ makes use of a number of sources. These sources must be understood in connection with $A$. We list some possible connections for the mentioned source types.

\begin{description}
\item{\em Real estate}. $A$ may own, lease or rent a collection of real estate items which are `used~by~$B$'.
\item{\em Vehicles}. $A$ may own, lease or rent vehicles, $A$ may have a contract with a provider of transportation services.
\item{\em Works of art and culture}. These are often owned and sometimes leased.
\item{\em Tools and equipment}. $A$ may own, lease or rent tools and equipment, $A$ may have a contract with a provider of technical services.
\item{\em Control codes}. $A$ may own a control code under some license; $A$ may have a temporary right of usage under some license;
$A$ may own a control code together with some or all underlying IPR (e.g., copyright on source code, copyright on requirements specifications, design documents, patents on architectures and designs, patents on control code production methods).
\item{\em Information bases}. $A$ may own the right to maintain the information based on formalized agreements with parties who have
provided these data. $A$ may own the right to use the information for a specified time frame and for specified purposes and with
a limited degree of usage (concurrent users).
\item{\em Employees and relationships}. $A$ may have persons as employees with contracts that prevent their employment with
(specified) other units. $A$ may have contracted persons not employed by other means than standardized employment. $A$ may have
agreements with other parties who promise the cooperation of their expert employees for certain tasks that $A$ expects $B$ to perform.
\item{\em IPR}. IPR can be owned, licenses can be obtained for limited time and purposes.
\item{\em Knowledge}. $A$ may have exclusive access to external expert knowledge. $A$ may have secured proprietary knowledge in contracts with employees, customers and suppliers.
\item{\em Contracts and goodwill}. Contracts are part of $A$'s legal position; goodwill may be connected to brand names (which can be owned).
\item{\em Finance}. Financial sources can be understood as a part of $A$'s contractual position. Financial sources can be linked to $A$ in
a variety of ways which may be all subsumed under `being available to'.
\end{description}

\subsubsection{A ranking of titles to sources from the outsourcer's perspective}\label{titleranking}
Titles to sources can be ranked on a scale which measures a degree of ownership. This can be done from the perspective of the outsourcer and also from the perspective of the insourcer. These rankings are inverse of one another. We consider three types of sources mentioned above, leaving a consideration the other eleven types to the reader's imagination.\\

\noindent From an outsourcer's ($A$) perspective we have the following decreasing sequences of titles for source types Person, Tool, Equipment and IPR:

\begin{description}
\item{\em Personnel.}
Because in many cases personnel constitutes the most important source the listing begins with a catalogue of `titles' on personnel. Moving downward in this list of titles (given a certain resource) is the direction of outsourcing for $A$ and the direction of insourcing for $X$. The ordering of titles ranks usage more important than ownership; this may be less convincing in some specific case.

\indent The list of titles are symmetric for $A$ and $X$. The first half of titles is considered positive for $A$ (and negative for $X$), indicating that a
source belongs to $A$ to some (positive) extent, the second half of the titles is negative for $A$ and positive for $X$. Below, unit $C$ is a contractor different from $A$, $B$, $X$ and $Y$.

\begin{enumerate}
\item $P$ is a permanent employee of $A$ in its subunit $B$;
\item $P$ is a temporary employee of $A$ in its subunit $B$;
\item $P$ has been contracted by $A$ from $C$ for a project carried out by or for subunit~$B$;
\item $P$ has been contracted by $A$ from $X$  for a project carried out by or for subunit~$B$;
\item $P$ has been contracted by $X$ from $A$ for a project carried out by or for subunit~$Y$;
\item $P$ has been contracted by $X$ from $C$ for a project carried out by or for subunit~$Y$;
\item $P$ is a temporary employee of $X$ in its subunit $Y$;
\item $P$ is a permanent employee of $X$ in its subunit $Y$.
\end{enumerate}

\item{\em Tools and equipment.} For a tool or a piece of equipment titles occur in this order:
\begin{enumerate}
\item $H$ is owned by $A$ and used by personnel of $A$ for work in its subunit $B$;
\item $H$ is leased or rented by $A$ from $C$ and used by personnel of $A$ for work in its subunit $B$;
\item $H$ is leased or rented by $A$ from $X$ and used by personnel of $A$ for work in its subunit $B$;
\item $H$ is leased or rented by $X$ from $A$ and used by personnel of $X$ for work in its subunit $Y$;
\item $H$ is leased or rented by $X$ from $C$ and used by personnel of $X$ for work in its subunit $Y$;
\item $H$ is owned by $X$ and used by personnel of $X$ for work in its subunit $Y$.
\end{enumerate}

\item{\em IPR.} For an object L subject to IPR regulations (say copyright), the following order of titles is plausible:
\begin{enumerate}
\item $L$ is owned (including copyrights) by $A$ and used by personnel of $A$ for work in its subunit $B$;
\item $L$ is licensed by $A$ from $C$ and used by personnel of $A$ for work in its subunit $B$;
\item $L$ is licensed by $A$ from $X$ and used by personnel of $A$ for work in its subunit $B$;
\item $L$ is licensed by $X$ from $A$ and used by personnel of $X$ for work in its subunit $Y$;
\item $L$ is licensed by $X$ from $C$ and used by personnel of $X$ for work in its subunit $Y$;
\item $L$ is owned (including copyrights) by $X$ and used by personnel of $X$ for work in its subunit $Y$.
\end{enumerate}
\end{description}

\noindent Titles can be weighted in terms of their financial impact or in terms of some other degree of importance. No such attribution of weights can be designed in advance and be valid for all practical cases. But in a concrete setting and with a specific unit $A$ and its subunit $B$ at hand, even if no $X$ and $Y$ have yet been identified (provider/insourcer selection has not yet taken place), it is probably doable to assign a weight to each title so that portfolios of titles can be compared by summing the weight of the title to each of the sources. We'll investigate qualitative and quantitative assessments of sourcing equilibria further in Section~\ref{qualiquanti}.

\subsection{The outsourcing transformation and its precondition}\label{outsourcingTransformation}
Outsourcing is about: some unit $A$, a subunit $B$ and sources made use of by $B$.  At this stage it is assumed that $A$ has an {\em identity} (as formulated in its mission statement)  which does not critically depend on any of the legal titles to sources that its subunit $B$ makes use of. To illustrate this, consider a museum $A$ that has been set up to show and maintain certain specific works of art in its division $B$. These works are sources, but $A$'s identity cannot be decoupled from such sources.

Precisely this condition, or rather {\em precondition} when taking the order in time in account, ensures that:

\begin{quote}
$A$'s identity can be preserved when titles to $B$'s sources are changed.
\end{quote}

\noindent This condition necessitates the distinction between $A$ and $B$ when working towards a specification of outsourcing.\\

\noindent An {\em outsourcing transformation} for sources of $B$ (from the perspective of the outsourcer $A$) comprises a package of modifications to $A$'s organization and structure which can be understood as a partial or complete replacement of $B$ by some other unit -- earlier denoted as $B_R$, the retained unit derived from $B$ -- which either delivers or guarantees (manages) the same services to sister units of $B$ in $A$ and to the external customers of $A$ that were served by $B$.

In the case of insourcing an entirely symmetric setting may be assumed. However, if $A$ is outsourcing it may be the case that some other (or new) unit $Y$ is changing so much that it is not valid to describe $Y$'s modification as an insourcing transformation. Indeed, if $Y$ is newly created in order to perform outsourcing sources from $B$ (in $A$) then $Y$ is not insourcing anything. The situation is symmetric: if an insourcer deals with $B$ rather than $A$ then it may insource $B$'s sources but there is no outsourcer whose identity is left essentially constant.\footnote{Indeed, $B$'s identity is not supposed to be independent from its family of titles to the sources it uses.}

Assuming that $A$ and $X$ exchange sources for their respective subunits $B$ and $Y$. Then after the exchange $B$ and $Y$ have been transformed to $B_R$ and $Y_R$ (for symmetry reasons the same notation is used). In the simplest case both $B$ and $Y$ keep their identity constant and $B_R$ and $Y_R$ working concurrently together deliver the same bundle of services as before.

\subsection{A postcondition for outsourcing}\label{postcondition}
Complementary to the precondition which is supposed to apply before an outsourcing transformation, a {\em postcondition} captures some requirements on the state after outsourcing. In this situation $A$ is said to perform outsourcing of sources of $A$'s unit $B$ to $X$ (and $X$ is said to insource sources for $X$'s subunit~$Y$) if:
\begin{enumerate}
\item the family of $A$'s titles to (non-financial) sources of $B_R$ is substantially smaller (in terms of a weighted sum of rankings of titles and preferably as well in terms of financial valuations) than the original family of titles to sources of $B$, and;
\item $A$'s titles to financial sources may have decreased, while $X$'s titles to financial sources may have increased;\footnote{In other
words: $X$ pays $A$ for the transfer of titles to sources as well as for investments (acquisition of additional sources) needed to
provide the services required to realize service invariance.}
\item the family of titles to (non-financial) sources of  $Y_R$ is larger (in terms of a weighted sum of rankings of titles and
probably in terms of financial valuations) than the original family of titles to sources of $B$.
\end{enumerate}

\noindent One may wonder whether one should demand as part of the postcondition that an identifiable residu of $A$ should remain. We think we should not. With this point of view we can come back to the remark made in the beginning of this section (page~\pageref{nconcepts}). There we said that it would appear that backsourcing can't always be used to indicate the reversal of an insourcing. Here we can explain why.

The concept of backsourcing indicates an asymmetry between insourcing and outsourcing because it is performed, if at all, on the initiative of the outsourcing unit. If no outsourcing unit is left because the insourcing which takes place is a full unit incorporation backsourcing cannot take place for the obvious reason that there is no unit to which the sources may return. So, backsourcing is by default a transformation that is better described as {\em reverse-outsourcing}. If reverse-outsourcing is not possible because the outsourcer has evaporated then a transformation in which the insourcer has the initiative could in principle still be enabled. Following the same naming scheme one could call this: {\em reverse-insourcing}. 

With reverse-insourcing a new unit is created that functionally matches the former outsourcer. Of course, this calls for a historical reference model that might not be available. So, the functionality of the created unit may differ considerably. In that case the transformation takes us outside the sourcing realm and into the wider area of organizational transformations.

\subsection{Outsourcing progression}\label{progression}
An {\em outsourcing progression} is a progression which consists of a number of steps that lead from one sourcing equilibrium to another sourcing equilibrium and which transforms the sourcing equilibrium in such a way that a match with the precondition and the postcondition of an outsourcing transformation can be obtained. In the absence of such a match a progression still changes the sourcing equilibrium and may be understood as an outsourcing transformation, though not necessarily one of a known kind.

An outsourcing progression takes place concurrently with the residual activity of units $A$ and $X$. As said, the progression itself takes the form of a sequence of steps each of which can be understood as a modification of an outsourcing equilibrium. At this stage we assume that the extent of the notion of an outsourcing progression step can be inferred from the description of an outsourcing equilibrium that has been provided.

It should be noticed that although changing the legal title to some source may be considered a step, it is plausible that in practice it will consist of a thread of steps which may itself require project management and planning. Even to the point were an outsourcing progression results from the execution of an outsourcing plan that makes use of multi-threaded concurrency. Such a plan will be based on estimates for $A$ and $X$ concerning their respective advantages. Such estimates are made in business plans and according to Guus Delen both $A$ and $X$ need to understand each other's business case for the plan to be valid \cite{Delen2005}.

\subsection{Survey of terminology and some derived terminology}\label{termsurvey}
In conclusion, we understand outsourcing more restricted than usual by only taking the transformational meaning into account, while we are more liberal by not requiring the presence of a target service provision contract, nor any other guarantee that the outsourcing transformation will not damage $A$'s operations. The mere outplacement of sources in combination with the insourcer's ability to compensate for $B$'s loss of ability (after having become $B_R$) by providing $B$'s original services to its sister units in $A$ as well as perhaps to some clients of $A$ suffices for the qualification of the sourcing equilibrium as result of an outsourcing transformation.

Below we sum-up the terminology introduced above and complement it with some additional terminology (in {\em italics}). When an outsourcing transformation between $A$ and $X$ takes place as specified above the following terms and phrases
are in order:
\begin{enumerate}
\item $A$ is the outsourcing unit.
\item $B$ is the outsourcing subunit. It must be sufficiently small, relative to $A$, to allow $A$'s identity to be preserved when its titles to sources are terminated. And it must be sufficiently large to contain all sources which have their titles changed in the transformation at hand.
\item The service portfolio of $B$ together with that of $Y$ is the {\em invariant service portfolio} for the outsourcing progression.\footnote{It may
be the case that $Y$, a subunit of $X$, needs to divest some of its services (constituting a divesting transformation involving only $X$)
before the outsourcing progression takes place in order to be able to keep the combined service portfolios of $B$ and $Y$ invariant.}
\item The services from the invariant service portfolio which were provided by $B$ to sister units within $A$ or to clients of $A$ and which
are provided by $Y_R$ after the outsourcing progression took place are referred to as the {\em transferred services}.
\item Sources of $B$ for which the title disappears or decreases (say from ownership to rent or to license) are {\em unsourced}, these are also called {\em abandoned sources}.
\item Sources of $B$ for which the title is transferred from $B$ to $Y$ are said to be {\em properly outsourced}.
\item If there is a plan for insourcing the properly outsourced sources together with a re-acquisition plan of unsourced sources in order to reconstruct $B$'s original functionality after some time, the outsourcing progression is {\em technically reversible}.
\item If in addition to the outsourcing being technically reversible, there is a contract of $A$ with $X$ which permits $A$ at some stage to terminate the use of services which $Y_R$ provides to it (since the outsourcing took place) and also to properly insource (a large part of) the previously properly outsourced sources then the outsourcing progression is said to be {\em contractually reversible}.
\item If an outsourcing progression is reversed (it need not have been contractually agreed that this might occur) then we speak of a {\em backsourcing progression}. A backsourcing progression is an outsourcing progression for $X$ and an insourcing progression for $A$. However, if the original outsourcing  was contractually reversible, backsourcing can take place even if $A$ cannot see a positive business case for $B$, unless it takes the penalty for $B$'s non-compliance with the reversibility item from its original insourcing contract into account.
\item $X$ is the insourcing unit.
\item $Y$ is the insourcing subunit.
\item If $Y_R$ is providing services to $A$ which were not delivered by $X$, that is if $X$ has increased its volume of services provided to $A$,
then $X$ is called a {\em provider} (relative to the out/insourcing transaction).
\item Sources of $Y_R$ which (titles) have neither been inherited from $Y$ nor transferred from $B$ are {\em acquired sources}, also called {\em additional sources}.
\item Sources of $Y_R$ for which the title has been transferred from $B$ to $Y$ are said to be {\em properly insourced}.\footnote{Proper insourcing of sources
 has many variations because an ownership of source $S$ by $A$ may turn into an ownership of $S$ by $X$ but also into an ownership by
 yet another unit $Z$ in combination with a lease of $S$ by $X$ from $Z$ and so on.}
\item If $X$ agrees by contract to provide services $S$ through $Y_R$ (with help of $B_R$) to other units of $A$ and clients of $A$ then
this contract is called the {\em target service provision contract}.\footnote{If $A$ takes the initiative for outsourcing then its objective is to sign a target service provision contract at minimal cost but with good quality of service.}
\item In the presence of a target service contract, $A$'s activities concerning the monitoring and control of the fulfillment of the target service provision contract are called {\em client side contract management}. $X$'s activities concerning the same contract are referred to as {\em provider side contact management}.
\item In the presence of a target service provision contract and after its expiration the sourcing equilibrium may change in the following ways:
\begin{description}
\item{\em Backsourcing.} If the outsourcing was contractually reversible (for $A$) then $A$ may choose to seek for a backsourcing transformation. As argued in Section \ref{postcondition}, a more systematic name for this activity is {\em reverse-outsourcing}.
\item{\em Outsourcing prolongation.} If $A$ was satisfied with the services that $X$ and $Y_R$ provide then a prolongation of the target outsourcing contract may be contemplated. This could be called the {\em trivial outsourcing transformation} a transformation in which the sourcing equilibrium is left unchanged.
\item {\em Reverse-insourcing.} This may be contractually guaranteed in the case that the out/in\-sourcing transformation has entirely removed the unit from which sources were outsourced. This situation leaves the initiative for a reversal with the insourcer.
\item {\em Reverse-outsourcing.} See `backsourcing' above.
\item{\em Follow-up sourcing to a third party.} If $A$ is less satisfied with $X$'s performance
follow-up sourcing to another party, say $Z$ may be considered by $A$. In follow-up sourcing $A$ may need
contractually regulated consensus with $X$ that some of the assets which it has acquired as a result of the outsourcing deal,
can be further transferred to the third party $Z$.\footnote{Outsourcing prolongation is a simple case of follow-up sourcing. Backsourcing is
potentially more complex than follow-up sourcing to a third party because $A$ may need to introduce new additional sources for $B_R$
(or rather $B_{RR}$, the incarnation of $B_R$ after backsourcing) under the assumption that to implement
backsourcing an attempt is made to reconstruct the original sourcing equilibrium.}
\item{\em Source and service reconstruction.} If none of the above applies and $A$ needs to maintain its part (the part of which it is a client)
of the invariant service portfolio available $A$ may re-acquire sources (or acquire comparable but different sources)
that have been abandoned or transferred to $X$ in the original outsourcing progression, in order to re-establish the service provision capability
for the transferred services. The terminology is more liberal than backsourcing in the sense that during source and service reconstruction $A$ need not have $B_R$ in the role of the insourcing subunit.
\item{\em Multiple follow-up sourcing.} $A$ finds different service providers (all different from $X$)  for follow-up provision of
distinct parts of the transferred services.
\item{\em Heterogeneous multiple follow-up sourcing.} $A$ combines the mechanisms above for distinct parts of the transferred services.
\end{description}
\end{enumerate}

\section[Quantification and qualification of sourcing equilibria]{Quantification and qualification of sourcing\\equilibria}\label{qualiquanti}
From a rational business perspective it is expected that outsourcing and insourcing take place because assessment of a sourcing equilibrium reveals the likelihood that a better arrangement can be achieved. Of course, many motives are conceivable, for instance, appealing management trends, hunches, perceived but false opportunities.

With the terminology from the previous section we can now look at assessing a given sourcing equilibrium. We contemplate, again, a unit $A$ and its subunit $B$ which may become involved in outsourcing sources used by $B$ to subunit $Y$ of unit~$X$.

\subsection{Cost estimation of a source portfolio}
In a sourcing equilibrium state preceding outsourcing $B$ makes use of a portfolio of sources and for each source $A$ has some legal title for the use of that source which is likely to fit the above listing of source type specific legal connections. The family of source titles that $B$ makes use of as well as $B$'s dependency of $A$ can be quantified in terms of costs as follows.
\begin{enumerate}
\item Capital costs of owned sources.
\item Operational costs of owned, leased, rented or licensed sources.
\item Costs of lease, rent and licenses.
\item Costs of employees and non-employed personnel. (Minus the advantage that lies in their not having been employed or
contracted by $A$'s competitors in the market of $B$'s services.)
\item Insurance policies for owned and leased sources.
\item The cost of $A$'s management efforts needed to survey and control $B$'s activities inside $A$.
\item Obligations for $A$ to accept modifications of its sourcing structure due to previous outsourcing transformations.\footnote{This aspect is needed if backsourcing is to be classified  as an outsourcing transformation, in combination with the requirement that both parties understand each others business cases.}
\end{enumerate}

\noindent Quantification and assessment of costs of outsourcing to support management decisions have been studied by Chris Verhoef in a theoretical as well as a applicable manner \cite{Verhoef-outsourcingdeals-2005}.

\subsection{Sourcing architecture}
A {\em sourcing architecture} is a normative description of a sourcing equilibrium. Assuming that architecture precedes implementation and that an implementation achieves what is intended we will not make a difference between sourcing equilibrium and sourcing
architecture.\footnote{Interface groups as defined in \cite{BergstraPonse2007} have been developed for the description of sourcing architectures and in particular for the description of unit interfaces that supports their modular composition.}

An outsourcing plan will contain a sourcing architecture for the original composition of $A$ which may have been found by means of reverse architecting (architecture extraction). In addition it will contain a sourcing architecture for the new form of subunit $B$ of $A$ and also for the new form of the subunit $Y$ of $X$.

\subsection{Internal and external source}
A source $S$ used by $B$ is {\em internal} from the perspective of $A$ if its title on $S$ is positive (see the ranking in Section \ref{titleranking}) compared to the titles on $S$ by any other unit that makes use of the services which $B$ creates by means of source $S$. A source $S$ used by $B$ is {\em external} (from the perspective of $A$) if it is not internal.

It is plausible that after outsourcing from $A$ its collection (weighted sum) of internal sources has decreased and its collection of external sources has increased.

\subsubsection{Relative degree of internal and external sourcing}
For $A$ and $B$ and some source type $T$ it may be possible to determine a benchmark by inspecting other (competing) units operating in a similar market condition. The benchmark indicates what proportion of sources of type $T$ used by $B$ are normally external to $A$. Such proportions can be related to performance metrics in order to determine a best shape for the sourcing equilibrium under the assumption that benchmarking with comparable units is valid. A relative degree of external sourcing can be derived directly from the relative degree of internal sourcing.

\subsubsection{Absolute degree of internal and external sourcing}
If all activities of $B$ have been financially quantified it may even be possible to determine an objective proportion of type-$T$ sources which are internal, thus producing a measure of internal sourcing which is independent of the sourcing equilibrium of competing units. Such a measure can be used to analyze the long term evolution of a unit. An absolute degree of external sourcing can be derived from the absolute degree of internal sourcing.

It is plausible that after outsourcing from $A$ its absolute degree of internal sourcing has decreased and its absolute degree of external or external sourcing has increased.

\subsection{Internal and external service provisioning}
A service used by some unit of $A$ is internally provided by $B$ {\em to} $A$ if $B$ creates the service from its underlying sources. Some of these sources may be external to $A$, in which case some external unit provides access to the source via an appropriate service. That specific service is external to $A$.

Symmetrically, a service used by another unit $C$ and created by subunit $B$ of $A$ is an external service {\em of} $A$. A service can be both external and internal at the same time.\footnote{For instance a restaurant catering mainly $A$'s personnel may also serve external customers, e.g., members of $C$'s personnel.} A service can be served  simultaneously to different concurrent users.

It is plausible that after outsourcing from $A$ its  volume of internal sourcing has decreased and its volume of external or external sourcings has increased.

\subsubsection{Relative and absolute degree of internal and external service provisioning}
Under the assumption that unit $A$ is comparable with a family of other units and major differences occur mainly in their sourcing architectures, one may compare corresponding services and determine a relative degree of internal service provision.

A relative degree of external service provision follows at once, as does an absolute degree of internal or external service provisioning.

\section{Outsourcing and insourcing defined}\label{outsourcingdefined}
As was probably already clear to the reader, we simply define outsourcing as an outsourcing progression. That is, a sequence of steps which (by definition) achieves an outsourcing transformation.

The details of the definition of an outsourcing progression have been spelled out above (\ref{progression} and \ref{termsurvey}) for the unit to unit case where complementary to the outsourcing unit there is an insourcing unit which exists before the progression begins. For the case that the outsourcing progression creates a new unit within the insourcing unit the definitions must be marginally adapted. Outsourcing can be ahead, in progress, or completed. It can be started and interrupted, it can be dormant, restarted, and successfully continued.

It may be considered a weakness of our definition that an outsourcing progression cannot be unsuccessfully terminated for the simple reason that it fails to qualify as an outsourcing progression in that case. We'll digress on this below.

\subsection{A first candidate for outsourcing definition maintenance}\label{firstcandidate}
If once considers it unacceptable that an outsourcing progression cannot terminate unsuccessfully a solution could possibly be found by the introduction of an outsourcing plan. Moreover, if a business case for an outsourcing transformation must include the cost of performing the outsourcing progression it is necessary to have access to a plan before putting the progression into action.

Such a plan consists of a (probably multi-threaded) specification of a number of threads of actions (steps) which, when successfully performed (that is: for each of the actions its precondition was found valid during plan execution), produce an outsourcing progression. In an exceptional case the plan execution may come to an unsuccessful halt while reporting that outsourcing has not been achieved.\\

\noindent In practice it is plausible that an outsourcing transformation in the vast majority of cases results from executing an outsourcing plan, but it cannot be excluded that an outsourcing transformation takes place as the unintended side-effect of a progression of steps which came about from executing plans with entirely different objectives.

Therefore, we prefer the viewpoint that an outsourcing progression always terminates successfully by definition. Thus an outsourcing progression is a completion of the execution of an outsourcing plan, whatever the outcome. So, we reject this suggestion for a definition upgrade and we refrain from redefining an outsourcing as the progression produced by executing an outsourcing plan.

\subsection{A second definitional revision}\label{secondcandidate}
Suppose one accepts nevertheless the revision of outsourcing just mentioned (and rejected) and declares it to be the successful execution of an outsourcing plan. In that case it is plausible to include the planning generation task itself in the definition of outsourcing.

If performing the outsourcing progression is put in the hands of an external consultant or operator, this agent will be asked to generate and run-time adapt the relevant plans. The problem with a revision of the definition of outsourcing in this fashion is that increasingly aspects are involved which are unspecific for achieving an outsourcing transformation.

So, to be consistent (having rejected both candidate upgrades of the outsourcing definition), we must refer to a consultant who will manage and supervise (control) an outsourcing as an agent who {\em generates an outsourcing plan and who readjusts this plan when necessary during its execution}. Following this approach, the specification of the precondition and postcondition of the outsourcing transformation (Section \ref{outsourcingTransformation} and \ref{postcondition}) need to adapted in order to guide the plan generating efforts of this consultant.

\subsection{Unit to unit insourcing defined}
Given the definition of unit to unit outsourcing, the complementary phenomenon of unit to unit insourcing is captured by precisely the same definition. The case where insourcing terminates the existence of the outsourcing unit requires some marginal adaptations comparable to the modifications needed when dealing with outsourcing to a newly created unit.

\subsection{Multiple outsourcing}
It is easy to see that outsourcing can take place to two or more units at the same time, especially if no other unit alone can successfully insource the sources which an outsourcer intends to outsource. Quite complex structures of service delivery and consumption may emerge if different insourcers work together and in particular if the family of insourcing units is made responsible for self-organization. Clearly a plurality of insourcing units will operate concurrently in some fashion.

\subsubsection{Concurrent existence and activity of units}
Given a sourcing equilibrium different units operate concurrently. There seems to be nothing of any generality that can be stated about the concurrent activity of different units before and after outsourcing. A formal model of concurrency might be used to provide an imaginative definition of the concurrent activity of two or more units. A very simple formal model of concurrent activity is thread algebra as presented in \cite{BergstraMiddelburg2007}. In that model units produce threads which represent their sequential behavior and different threads are strategically interleaved (that is: interleaved by means of some plausible strategy).

After an outsourcing progression has been completed (that is the sequence of steps which constitute the outsourcing transformation in a particular case), outsourcing unit $A$ and insourcing unit $X$ must operate concurrently. This can be understood in terms of multi-threading if it is assumed that units (or rather subunits) have some form of centralized control which makes it plausible to imagine their activity as the run of a sequential program.

\subsubsection{Interaction between units}
For a description of inter-unit cooperation by means of multi-threading streams of artifacts and data moving between various units need to be formalized. Technically such formalizations can be taken from any theoretical model of concurrent computation. In the setting of outsourcing it seems reasonable to work with a global state space.

\section{Conclusions}\label{conclusie}
We are confident that we have provided a fairly detailed description of unit to unit outsourcing and insourcing which qualifies as an imaginative definition following~\cite{Bergstra2010a}. Many decisions have been taken with the effect that the concept of outsourcing and insourcing conceived this way becomes less inclusive. We've called our approach to achieve a theory of outsourcing and insourcing business mereology.

At this stage we cannot provide a significant validation of our definition. In principle one may validate the definition by making a survey of acknowledged outsourcing cases and finding out to what extent our definition can be applied. An application then consists of recognizing a part of the activities involved which qualifies as an (unit to unit) outsourcing progression.

In a complex organization (unit)  insourcing progressions and outsourcing progressions take place concurrently concerning different subunits. An interesting topic for further investigation which might profit from the precision of the definitions which we have developed is the presence of interference or feature interaction between different out/insourcing progressions. See, e.g., \cite{KimblerBouma1995} for a survey of feature interaction theory.\\

\noindent Many other issues can be investigated on the basis of the work in this paper. We mention some directions for future work.
\begin{enumerate}
\item Developing a textual specification of a well-known kind of outsourcing transformation in significant detail. For instance, the outsourcing of desktop support to a shared service center working for a number of subunits of a given unit.
\item Providing a survey of known outsourcing progressions and finding a match between those cases and the elements of the definitions that have been presented above.
\item Describing the concurrent activity that takes place in outsourcing progressions that result in a multiple outsourcing transformation.
\item Complementing our definitions with the additional complexity of legal processes, seen from the perspective of a prospective outsourcer, that govern the selection of candidate insourcing units and the decision making that leads to selection of one of them.
\item Complementing our definitions with strategic aspects that have not been mentioned above, such as the intended partnership between outsourcing unit and insourcing unit, and the expectation that an insourcing unit will support innovations envisaged by the outsourcing unit even in the absence of a compelling contractual basis.
\item Analyzing the specification of transformations (of a sourcing equilibrium) which must be understood as concurrently executed instances of outsourcing and insourcing transformations between a number of units. The introduction of a shared services center is a relatively simple example of a transformation of this kind.
\item Specializing the imaginative definition of outsourcing to a specific area of business activity, for instance LPO (legal process outsourcing, see for instance \cite{Patterson2008}).
\item Identification of the structural, quantitative and qualitative properties of a network of units as it evolves by sourcing transformations.
\end{enumerate}

\end{document}